\documentclass[aps, twocolumn,noshowpacs,preprintnumbers,amsmath,amssymb]{revtex4}

\usepackage{verbatim}
\usepackage{wasysym}

\usepackage{graphicx}
\usepackage{dcolumn}
\usepackage{bm}
\usepackage{wrapfig}

\begin{document}

\title{Hindered rolling and friction anisotropy in supported carbon nanotubes}

\author{Marcel Lucas}
\affiliation{School of Physics, Georgia Institute of Technology, 837 State Street, Atlanta, Georgia 30332-0430, USA}
\author{Xiaohua Zhang}
\affiliation{International School for Advanced Studies (SISSA), and CNR-INFM Democritos, Via Beirut 4, 34014 Trieste, Italy}
\author{Ismael Palaci}
\affiliation{School of Physics, Georgia Institute of Technology, 837 State Street, Atlanta, Georgia 30332-0430, USA}
\author{Christian Klinke}
\affiliation{Institute of Physical Chemistry, University of Hamburg, 20146 Hamburg, Germany}
\author{Erio Tosatti}
\affiliation{International School for Advanced Studies (SISSA), and CNR-INFM Democritos, Via Beirut 4, 34014 Trieste, Italy}
\author{Elisa Riedo}
\affiliation{School of Physics, Georgia Institute of Technology, 837 State Street, Atlanta, Georgia 30332-0430, USA}
\email{elisa.riedo@physics.gatech.edu.}

\begin{abstract} 

Carbon nanotubes (CNTs) are well known for their exceptional thermal, mechanical and electrical properties. For many CNT applications it is of the foremost importance to know their frictional properties. However, very little is known about the frictional forces between an individual nanotube and a substrate or tip. Here, we present a combined theoretical and experimental study of the frictional forces encountered by a nanosize tip sliding on top of a supported multiwall CNT along a direction parallel or transverse to the CNT axis. Surprisingly, we find a higher friction coefficient in the transverse direction compared with the parallel direction. This behaviour is explained by a simulation showing that transverse friction elicits a soft 'hindered rolling' of the tube and a frictional dissipation that is absent, or partially absent for chiral CNTs, when the tip slides parallel to the CNT axis. Our findings can help in developing better strategies for large-scale CNT assembling and sorting on a surface.

\end{abstract}

\maketitle

Carbon nanotubes (CNTs) are well known for their exceptional thermal, mechanical and electrical properties \cite{1,2,3,4,5,6}. For many CNT applications it is of the foremost importance to know their frictional properties. However, very little is known about the frictional forces between an individual nanotube and a substrate or tip. Here, we present a combined theoretical and experimental study of the frictional forces encountered by a nanosize tip sliding on top of a supported multiwall
CNT along a direction parallel or transverse to the CNT axis. Surprisingly, we find a higher friction coefficient in the transverse direction compared with the parallel direction. This behaviour is explained by a simulation showing that transverse friction elicits a soft 'hindered rolling' of the tube and a frictional dissipation that is absent, or partially absent for chiral CNTs, when the tip slides parallel to the CNT axis. Our findings can help in developing better strategies for large-scale CNT assembling and sorting on a surface.

Frictional forces are responsible for an enormous variety of phenomena in our everyday life including playing the violin, dancing the tango and driving a car \cite{7}. The same wide spectrum of friction applications can be found in nanobiotechnology. In particular, the frictional forces between concentric CNTs have attracted considerable interest for CNT potential use in nanomechanical \cite{2,3} and nanoelectrical systems \cite{8,9}. Experiments and theory agree that the translational and rotational interlayer forces between suspended concentric nanotubes are extremely small, with values of the translational shear strength ranging between 0.2 and 0.04 MPa (refs 10, 11) and values of the rotational shear strength varying from 0.85 to 0.05 MPa (refs 12, 13). The frictional behaviour of a nanotube deposited on a surface can clearly be very different. Further effects need to be considered, such as the possibility of rolling and the role of the adhesion force between the nanotube and the substrate. Atomic force microscopy (AFM) is an excellent tool to manipulate objects and investigate forces at the nanoscale. Previous AFM studies in air have investigated the sliding of a CNT on a graphite surface by CNT end-on pushing with an AFM tip \cite{14}. More recently, a CNT tip attached to a conventional AFM probe was slid along a nanotube suspended over a trench \cite{15}. Both studies found for the carbon-carbon interface a shear stress of a few megapascals. No friction studies are available about the role of CNT radius, CNT structural properties, sliding velocity and especially sliding direction relative to the tube axis. As CNTs are made of concentrically rolled-up graphene sheets, they might be expected to yield an essentially isotropic nanofriction when they slide on a surface parallel (longitudinal sliding) or perpendicularly (transverse sliding) to their axis.

\begin{figure*}[htbp]
  \centering
  \includegraphics[width=0.98\textwidth]{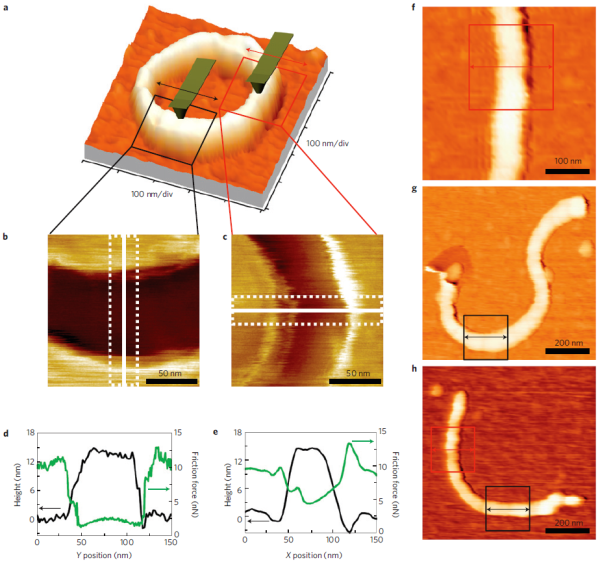}
  \caption{\textit{\textbf{Friction measurements on CNTs. a,} Topography image of a CNT of 7 nm radius. The fast scanning direction of the AFM tip is indicated by an
arrow ($X$ direction). \textbf{b,c,} Friction images of the highlighted longitudinal (\textbf{b}) and transverse (\textbf{c}) sections of the nanotube. \textbf{d,e,} Topography and friction force profile across the CNT. The topography profile (black solid line) is along the white solid lines in \textbf{b,c}. The friction force profile (red solid line) is the average profile inside the area delimited by the dotted line in \textbf{b,c}. The friction force profile in \textbf{d} is along the $Y$ direction and the one in \textbf{e} is along the $X$ direction. \textbf{f–h,} Topography images of three other CNTs of radius 11, 9 and 6 nm, respectively, where friction measurements were carried out inside the areas delimited by solid lines.}}
\end{figure*}

Here, we present a joint experimental and theoretical investigation of the frictional forces encountered by a nanosize tip sliding on top of a supported multiwall CNT along a direction parallel or transverse to its axis, as a function of normal load, sliding velocity and tube radius. Experimentally, the longitudinal shear strength is found to remain almost constant with varying CNT radius, whereas the transverse shear strength decreases with increasing radius, reaching the value of the longitudinal shear strength for radii larger than 10 nm. Experiments and theory find that for a fixed total normal force, that is, applied normal load plus adhesion force, the friction force, $F_{F}$, in the transverse direction is higher than in the longitudinal direction. In particular, the theoretical simulations find that the transverse-longitudinal friction anisotropy, $a = F_{F}^{Tran} / F_{F}^{Long}$, can be as large as 20 in non-chiral CNTs and about 2-3 in chiral CNTs. The experiments, carried out on multiwall CNTs with random chirality, show that a can be as large as 3. This unanticipated behaviour is explained by the theoretical simulations. We find that transverse sliding probes not only the elastic deformations of the stiff carbon-carbon bonds of the graphitic basal planes, common to the longitudinal sliding, but also a much softer overall swaying motion of the tube, akin to a hindered rolling. This added softness is the source of the extra transverse tip dissipation and increased friction.

\begin{figure}[htbp]
  \centering
  \includegraphics[width=0.45\textwidth]{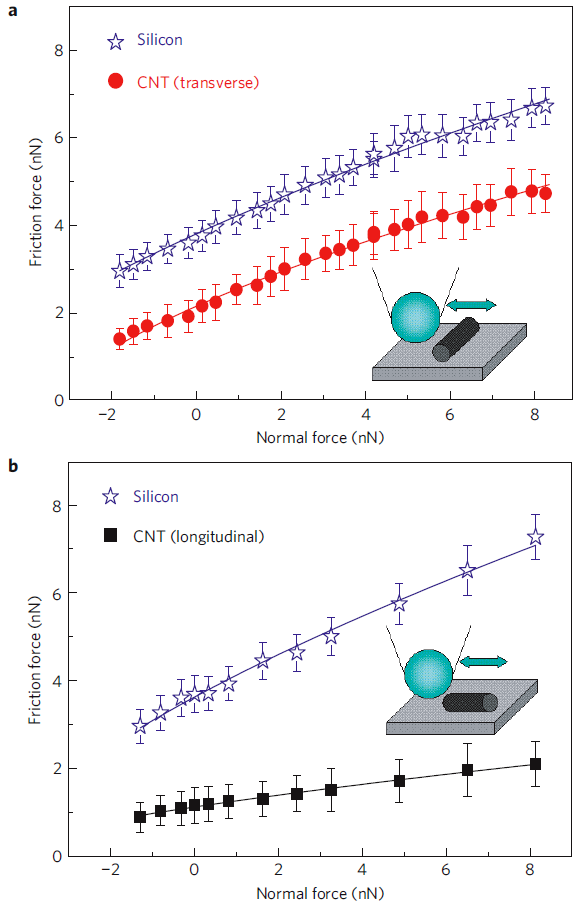}
  \caption{\textit{\textbf{Frictional forces for transverse and longitudinal sliding. a,b,} Frictional force as a function of the normal load for silicon and for
transverse (\textbf{a}) and longitudinal (\textbf{b}) sliding on top of a CNT, measured as described in Fig. 1. The CNT radius is $\sim 3.5$ nm in the longitudinal section and $\sim 5$ nm in the transverse section. The sliding velocity was 0.8 $\mu$m/s. The solid lines are fits to the data by using $F_{F} = \tilde{\mu} (F_{N} + F_{adh})^{2/3}$. The error of $F_{F}$ is obtained from the fluctuations in the values of friction in the investigated areas on top of the nanotubes (see Supplementary
Information for more details).}}
\end{figure}

The multiwall CNTs used in this study are produced by a catalytic chemical vapour deposition method and deposited on a flat silicon substrate. The structural properties of the CNTs were characterized by transmission electron microscopy, which indicated that these CNTs have several defects and a constant ratio of external to internal radii of $R_{NT} / R_{int} = 2.8 \pm 0.7$ (see Supplementary Fig. S4). The topography and the friction properties of the CNTs were characterized by AFM. To measure the transverse and longitudinal friction force on a nanotube, we looked for a statistical number of CNTs on the silicon surface and we took simultaneous topographical and frictional images in different locations of the nanotube and at different scales. In Fig. 1, we show some examples of CNTs that have been measured. We usually look for nanotubes that lie on the substrate with some sections parallel and some other sections perpendicular to the tip scan direction. We then zoom into the desired section and we acquire a friction and topography image (Fig. 1b,c). This method also has the advantage that for each friction/topography measurement, we acquire during the same scan FF of both the nanotube and the silicon substrate, which can be used as our reference system. Typical cross-sections of
the friction and topography images for longitudinal and transverse measurements are shown in Fig. 1d,e. Figure 2a,b shows a typical plot of $F_{F}$ acquired on top of a CNT as a function of the normal load, $F_{N}$, for transverse (Fig. 2a) and longitudinal (Fig. 2b) sliding. In each graph, we also add the results for the silicon substrate, which is the reference surface. As a first approximation, $F_{F}$ between the AFM tip and the silicon substrate or between the tip and the CNT,
could be expressed as:

\begin{equation}
  F_{F} = \mu (F_{N} + F_{adh})
\end{equation}

where $\mu$ is the friction coefficient and $F_{adh}$ is the adhesion force between the AFM tip and the investigated sample. However, it is well known that this linear relationship becomes a quasi-linear relationship at the nanoscale, where the "single contact" geometry is very likely. In fact, the friction force is proportional to the tip-sample contact area, $A(F_{N} + F_{adh})$:

\begin{equation}
  F_{F} = \sigma \cdot A (F_{N} + F_{adh})
\end{equation}

where the proportionality factor is called shear strength, $\sigma$ (ref. 16). The contact area between a spherical tip (AFM tip) and a cylinder (CNT) or a flat surface (silicon substrate) can be expressed as a function of $F_{N}$, $F_{adh}$, elastic moduli, tip radius, $R_{tip}$, and cylinder radius, $R_{NT}$, by using continuum mechanics theories, such as the Hertz theory \cite{17}. The details of the calculation of $A (F_{N} + F_{adh})$ are reported in the Supplementary Information. As a result, one obtains that $F_{F} = \tilde{\mu} (F_{N} + F_{adh})^{2/3}$. Depending on the nature of the adhesive forces, this equation can be slightly different; however, a discussion of the different contact mechanics models is beyond the scope of this letter and we refer the readers to refs 18 and 19 for details. Figure 2 seems indeed to indicate a 2/3 power-law dependence, but more importantly it clearly shows that the frictional forces are very different when the tip slides parallel or perpendicularly to the CNT long axis. The slope of the $F_{F}$ versus $F_{N}$ curve for the transverse CNT is similar to the slope measured on Si and it is about three times the slope measured on the longitudinal CNT. On the other hand, $F_{adh}$ remains almost the same for the two sliding directions (5.2 nN for longitudinal sliding and 3.3 nN for transverse sliding, as obtained by fitting the data with $F_{F} = \tilde{\mu} (F_{N} + F_{adh})^{2/3}$. No substantial differences have been detected in the $F_{F}$ versus $F_{N}$ curves acquired on the same CNT and for the same sliding direction but for different sliding velocities, in the range $\nu = 0.8 - 4$ $\mu$m/s.

\begin{figure}[htbp]
  \centering
  \includegraphics[width=0.45\textwidth]{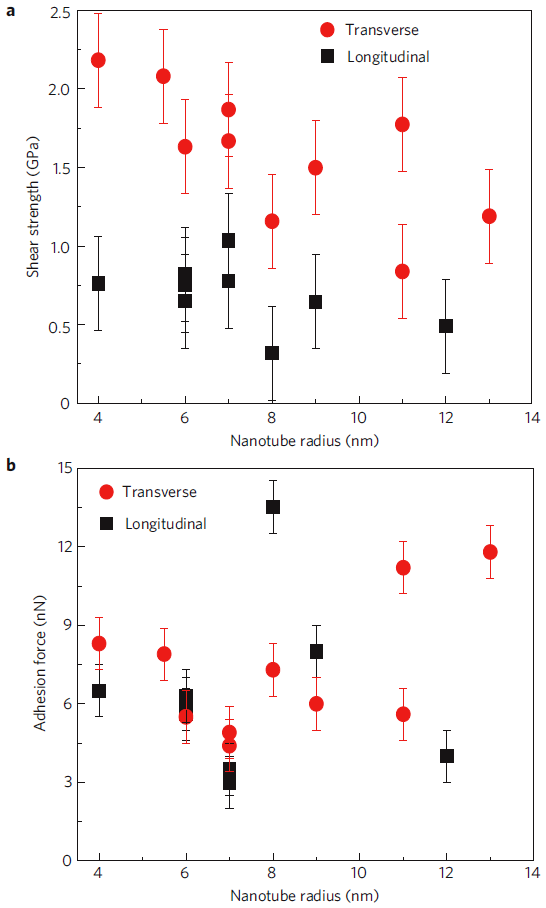}
  \caption{\textit{\textbf{Shear strength and adhesion force. a,b,} Shear strength (\textbf{a}) and adhesion force (\textbf{b}) for transverse and longitudinal sliding on top of a CNT as a function of the tube external radius. The sliding velocity was 2 $\mu$m/s for all of these measurements. The error on the shear strength is determined by fitting FF versus FN with equation (2), as described in the
Supplementary Information. The error on the adhesion force is obtained by
the fit of $F_{F}$ versus $F_{N}$ with equation (1).}}
\end{figure}

To further analyse the experimental data and compare the friction behaviour of CNTs with different radii, we derive from the experimental $F_{F}$ versus $F_{N}$ curves the shear strength, as introduced in equation (2) and $F_{adh}$ (see Supplementary Information and the Methods section). Figure 3a,b shows $\sigma$ and $F_{adh}$, respectively, as a function of the CNT radius for transverse and longitudinal sliding, at fixed velocity (2 $\mu$m/s). The longitudinal shear strength
($\sim$0.5 GPa) is independent of the tubes' radii in the range between 4 and 12 nm. This value is three/four orders of magnitude larger than $\sigma$ measured between the walls of the CNTs during telescope-pulling experiments. However, it is only two orders of magnitude larger than the values (about 4 MPa) found in experiments where a CNT slides on graphite or a CNT tip slides on another CNT in air. Furthermore, from our experiments carried out on highly ordered pyrolytic graphite in air with a Si tip we find $\sigma_{HOPG} = 3-60$ MPa, in good agreement with previous results \cite{20}. The presence of water, the tip-CNT contact geometry \cite{19} and the presence of defects in the CNT seem thus to be the reason for a large value of $\sigma$ measured during our experiments on CNTs. By comparing longitudinal and transverse sliding, the experiments indicate that in the transverse direction $\sigma$ is about three times larger than in the longitudinal
direction for a tube radius of 4 nm. For larger tubes, $\sigma$ slightly decreases in the transverse direction, with some fluctuations that could be related to a different CNT chirality and/or to the interaction of the CNT with the substrate, as suggested by the values of $\sigma$ for the radius of 11 nm for which $\sigma = 0.825$ GPa when $F_{adh} = 11.2$ nN and $\sigma = 1.781$ GPa when $F_{adh} = 5.6$ nN.

\begin{figure*}[htbp]
  \centering
  \includegraphics[width=0.98\textwidth]{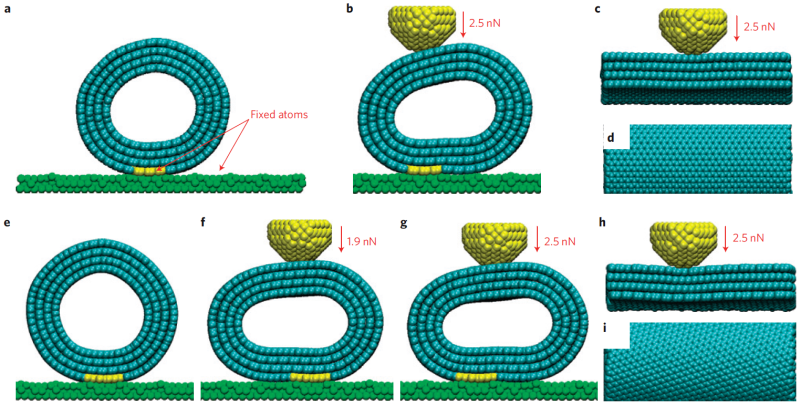}
  \caption{\textit{\textbf{Molecular dynamics simulation of the tip–nanotube interaction. a,b,} Non-chiral tube without tip (\textbf{a}) and under 2.5 nN normal force (\textbf{b}). \textbf{c}, The sectional view of the dimple. \textbf{d}, View of the outer tube (armchair). \textbf{e–g}, Chiral tube without tip (\textbf{e}) and under 1.9 nN (\textbf{f}) and 2.5 nN (\textbf{g}) with the tip pressing on two different spots. \textbf{h}, Sectional view, under 2.5 nN. \textbf{i}, Outer view of the chiral outer tube. Smaller sizes of both CNT and tip account for numerically smaller frictional forces compared with the data of Fig. 2.}}
\end{figure*}

To understand the observed frictional behaviour and the surprising transverse-longitudinal anisotropy, we conducted extensive molecular-dynamics simulations. Several model multiwall CNTs were built, with both armchair and chiral geometries and deposited on ideally flat crystalline substrates, the role of which is simply to stabilize the CNTs. The results presented here are for two different four-wall nanotubes, one fully armchair (26, 26)-(31,31)-(32, 40)-(41, 41) non-chiral tube and one (4, 46)-(32, 32)-(37, 37)-(18, 63) chiral tube. Well-tested force models (see the Methods section) were found to stabilize tubes with reasonable geometries and compliances (Fig. 4a,b). Each tube was approached by a nanosized diamond tip, controlled by a cantilever with three spring constants, exerting a variable load $F_{N}$ in the nanonewton range and laterally sliding back and forth with a speed $\nu \sim 2$ m/s along longitudinal and transverse directions relative to the tube axis. As in the experiments, the sliding amplitude was of the order of one nanometre and the frictional dissipation was extracted from the lateral force/displacement parallelogram area. The simulations clearly showed that the tubes deformed in two distinct manners during the tip pressing and sliding. First, we observe the formation of a dimple when the tip is on top of the tube centre (Fig. 4c,h). Second, we find a strong shape asymmetrization accompanied by an attempted lateral rolling of the tube (blocked in our case to prevent artificial runaway rolling caused by periodic boundary conditions) when the tip is off the tube centre (Fig. 4a,b,e-g). During motion, the longitudinal sliding simply displaced the dimple under the tip and along the tube. The transverse sliding instead gave rise to a further large deformation and lateral swaying of the tube. Under lateral pushing by the tip, the tube deformed as a whole, attempting a genuine hindered rolling back and forth in time. The calculated tip-nanotube frictional forces, determined by the dissipated energy over many cycles, are shown in Fig. 5, for both non-chiral and chiral nanotubes and for longitudinal and transverse motion. As observed during the experiments, the simulations show large transverse-longitudinal friction anisotropy. The larger transverse dissipation is clearly shown by simulation to be connected with the hindered rolling soft motion. Hindered by substrate adhesion, the tendency to roll decreases for increasing radii, consistent with the experimentally observed anisotropy decrease at larger radii. It should be noted that
there is a six-order-of-magnitude difference between the theoretical and the experimental sliding speeds (see Methods). However, in both cases the energy dissipation occurs by transfer of energy into the nanotube modes: smoothly in the simulation and by "kicks" during the probable stick and slip motion in the experiments.

\begin{figure*}[htbp]
  \centering
  \includegraphics[width=0.98\textwidth]{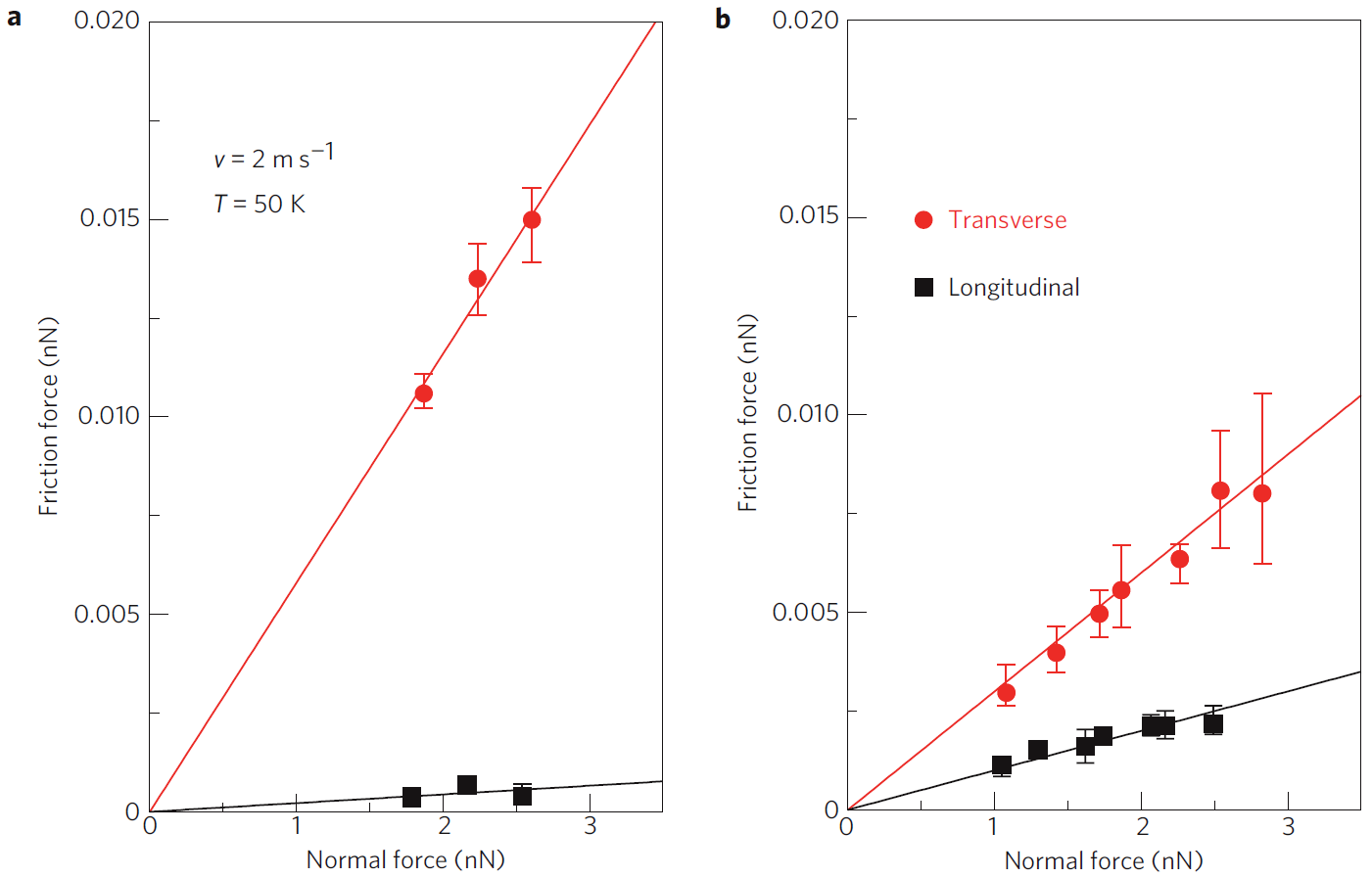}
  \caption{\textit{\textbf{Molecular dynamics simulation of tip–nanotube friction force versus normal load. a,}  Non-chiral nanotube (very large transverse–longitudinal anisotropy). \textbf{b}, Chiral nanotube (more realistic transverse–longitudinal anisotropy of about 2). The friction–load behaviour is to a good approximation linear (solid lines) and not a (2/3) power law as in the experiments, as the small size of the simulated tip and of the tip–CNT contact makes continuum contact mechanics invalid in this limit. Simulated friction extrapolates to zero at zero load, because the experimental zero-load situation corresponds to a finite load in the theory, the difference between the two corresponding to the experimental tip pull-off force, which measures adhesion, omitted in the simulation. The error bars indicate the friction fluctuations between different sliding loops.}}
\end{figure*}

Interestingly, the simulations show an anisotropy $F_{F}^{Tran} / F_{F}^{Long}$ as large as 20 in the fully armchair non-chiral tubes, but it drops to about 2 when the outer tube is chiral. Owing to chirality, the tip causes in a screw-like fashion some hindered rolling even for longitudinal sliding, owing to the absence of left-right symmetry. This suggests that our experimental multiwall CNTs were generally randomly chiral \cite{21}. It should be mentioned that structural defects
present in the CNT might also contribute in coupling the transverse and longitudinal dissipation modes in non-chiral CNTs, giving rise to an anisotropy coefficient smaller than 20. Dissipation due to motions outside the CNT and the occurrence of stick-slip probably also mitigates the frictional anisotropy. 

We have shown an anisotropy in the shear strength of CNTs in the transverse and longitudinal direction, which has its origin in the soft lateral distortion of the tube when the tip-tube contact is moving in the transverse direction. This friction anisotropy should be observable in very different geometries/situations, for example, when a CNT slides on a substrate parallel or perpendicularly to its axis. Our findings could help in developing better strategies for chirality sorting, large-scale self-assembling of surface-deposited nanotubes \cite{22}, design of CNT adhesives \cite{5} and nanotube-polymer composite materials \cite{6}. The conclusions of this letter more generally affect the ability to manipulate nano-objects on a surface.

\subsection*{Methods}

\textbf{AFM.} The topography and frictional force images were collected simultaneously using a Veeco CP-II AFM with a silicon tip of about 60 nm (PPP-CONT, NanoWorld) in air at room temperature, at a scanning speed ranging between 0.8 and 4 $\mu$m/s, for different normal force values. The normal force was between 8 and 2 nN initially and then decreased until the tip was out of contact with the sample. The cantilever normal spring constant of 0.11N/m was calibrated using the method of Sader and colleagues \cite{23}. The lateral force was calibrated using the wedge method on the Mikromasch TGG01 silicon grating \cite{24} (Supplementary Figs S1 and S2). The diameter of each section of CNT was inferred from the height of the imaged CNT lying on the silicon substrate. The contact area between the AFM tip and the silicon substrate or between the tip and a CNT was evaluated using the modified Hertz theory describing, respectively, the contact between a spherical tip and a planar substrate, or the contact between a spherical tip and a cylindrical tube \cite{16,25}.

To evaluate $F_{F}$ on the very top of the nanotube, after imaging the friction and topography of the desired section of the CNT and the surrounding silicon substrate, we zoom in on an area at the top of the tube where the topography and the friction look flat. For example, for a CNT of radius 7 nm, we zoomed in on an
area of 20 nm $\times$ 100 nm in the centre of the tube. By considering the convolution of the AFM tip (radius 60 nm) with a CNT of 7 nm, a width of $w_{convoluted} = \pm 10$ nm at the top of the imaged tube corresponds roughly to a width $w_{real} = \pm 1$ nm on top of the real nanotube. See Supplementary Information for more details on the evaluation of the contact area, friction measurements and error analysis (Supplementary Figs S5 and S6).

\

\textbf{Simulations.} In the molecular-dynamics simulations, CNTs were deposited on crystalline substrates, assumed for definiteness to be gold (111) (in place of the
real and more complex oxidized Si surfaces). The outer/inner tube radius ratios were substantially smaller in our simulations than in the experiments, implying first of all numerically smaller frictional forces and also a greater softness and a quantitatively larger tip-induced deformation from the cylindrical shape, as shown
in Fig. 4. Periodic boundary conditions were assumed parallel and perpendicular to the tube. For the substrate-nanotube interaction, a Lennard-Jones Au-C potential \cite{26} was used with a strength decreased by a factor of 10, an interaction that does not excessively alter the CNT cylindrical shape; all substrate atomic positions were assumed to be rigid, justified by the much larger substrate rigidity relative to the CNTs. Standard empirical Brenner potentials \cite{27} and Kolmogorov-Crespi potentials \cite{28} were used among CNT atoms. On the basis of the justification that the repulsive forces are short ranged and primarily responsible for the normal load, the tip-tube potential was assumed to be equal to the repulsive part of the Lennard-Jones potential, $V(r) = 4 \epsilon (\alpha /r)$ (ref. 12), with $\epsilon = 0.004$ eV and $\alpha = 0.328$ nm. A Berendsen thermostat was applied to the unconstrained atoms of the outer tube. The temperature was set at $T = 50$ K, different from the room temperature at which experiments were carried out, to avoid large fluctuations of the frictional force related to the small CNT sizes. Artificial massive lateral rolling of the tube, caused by the infinite tip replicas implied by the periodic boundary conditions and prevented in experiments by resistance to tip pushing by the strong substrate adhesion, was blocked in the simulation by freezing the relative carbon-substrate positions for a minimal number of outer tube strands in closest contact with the substrate (yellow coloured in Fig. 4). The rolling can be both leftwards and rightwards (Fig. 4f,g). The Young modulus ratio of silicon to CNT is roughly 170:30, meaning that the (silicon) tip and substrate deformations, although surely finite, are negligible compared with those of the CNT; thus, we chose a 525-atom rigid diamond tip to slide on the tube surface. The tip is connected to three supports ($X$, $Y$ and $Z$) by springs with constants $k_{x} = 50$ N/m along the sliding direction, $k_{y} = 500$ N/m along the other direction in the plane of the substrate and $k_{z} = 2$ N/m along the normal direction. The frictional force was calculated from the lateral force/displacement parallelogram area (see Supplementary Fig. S7). The speed of the support $X$ was set to be $\left| \nu \right| = 2$ m/s, sweeping forth and back, and $Y$ and $Z$ move together with the tip. This tip speed is many orders of magnitude larger than in the experiments, and in fact the simulated friction is smooth without stick and slip, the latter probably occurring in the experiments. Despite that, we do not expect the difference to affect the validity of our mechanism for the frictional anisotropy, because the CNT hardness against longitudinal shear and its softness against transverse shear found in the simulation should affect both frictional regimes in a similar manner.

\subsection*{ACKNOWLEDGEMENTS}

M.L. was financially supported by the Office of Basic Energy Sciences of the DOE (DE-FG02-06ER46293). E.R. acknowledges the NSF (DMR-0120967 and DMR-0706031) and DOE (DE-FG02-06ER46293) for summer salary support. Work in Trieste was supported by CNR under EUROCORES/FANAS/AFRI, as well as by a PRIN/COFIN contract.


\clearpage


\begin{thebibliography}{00}


\bibitem{1} Avouris, Ph. et al., IEEE International Electron Devices Meeting 2004, Tech. Dig. 525-529 (2004).
\bibitem{2} Cummings, J. \& Zettl, A., Science 289, 602-604 (2000).
\bibitem{3} Fennimore, A. M. et al., Nature 424, 408-410 (2003).
\bibitem{4} Klinke, C., Hannon, J. B., Afzali, A. \& Avouris, P., Nano Lett. 6, 906-910 (2006). 
\bibitem{5} Qu, L., Dai, L., Stone, M., Xia, Z. \& Wang, Z. L., Science 322, 238-242 (2008).
\bibitem{6} Vigolo, B., Poulin, P., Lucas, M., Launois, P. \& Bernier, P., Appl. Phys. Lett. 81, 1210-1212 (2002).
\bibitem{7} Persson, B. N. J., Tartaglino, U., Albohr, O. \& Tosatti, E., Nature Mater. 3, 882-885 (2004).
\bibitem{8} Cummings, J. \& Zettl, A., Phys. Rev. Lett. 93, 086801 (2004).
\bibitem{9} Forro, L., Science 289, 560-561 (2000).
\bibitem{10} Yu, M. F., Yakobson, B. I. \& Ruoff, R. S., J. Phys. Chem. B 104, 8764-8767 (2000).
\bibitem{11} Kis, A., Jensen, K., Aloni, S., Mickelson, W. \& Zettl, A., Phys. Rev. Lett. 97, 025501 (2006).
\bibitem{12} Bourlon, B., Glattli, D. C., Miko, C., Forro, L. \& Bachtold, A., Nano Lett. 4, 709-712 (2004).
\bibitem{13} Servantie, J. \& Gaspard, P., Phys. Rev. Lett. 97, 186106 (2006).
\bibitem{14} Falvo, M. R. et al., Nature 397, 236-238 (1999).
\bibitem{15} Bhushan, B., Ling, X., Jungen, A. \& Hierold, C., Phys. Rev. B 77, 165428 (2008).
\bibitem{16} Palaci, I., Fedrigo, S., Brune, H., Klinke, C. \& Riedo, E., Phys. Rev. Lett. 94, 175502 (2005).
\bibitem{17} Johnson, K. L. Contact Mechanics (Cambridge Univ. Press, 1987).
\bibitem{18} Schwarz, U. D., Zworner, O., Koster, P. \& Wiesendanger, R., Phys. Rev. B 56, 6987-6996 (1997). 
\bibitem{19} Carpick, R. W., Ogletree, D. F. \& Salmeron, M., Appl. Phys. Lett. 70, 1548-1550 (1997). 
\bibitem{20} Ritter, C., Heyde, M., Stegemann, B., Rademann, K. \& Schwarz, U. D., Phys. Rev. B 71, 085405 (2005). 
\bibitem{21} Hirahara, K. et al., Phys. Rev. B 73, 195420 (2006).  
\bibitem{22} Geblinger, N., Ismach, A. \& Joselevich, E., Nature Nanotech. 3, 195-200 (2008).
\bibitem{23} Sader, J. E., Chon, J. W. M. \& Mulvaney, P., Rev. Sci. Instrum. 70, 3967-3969 (1999).
\bibitem{24} Ogletree, D. F., Carpick, R. W. \& Salmeron, M., Rev. Sci. Instrum. 67, 3298-3306 (1996).
\bibitem{25} Arthur, P. \& Boresi, O. M. S., Advanced Mechanics of Materials 2nd edn (Wiley, 1986).
\bibitem{26} Luedtke, W. D. \& Landman, U., Phys. Rev. Lett. 82, 3835-3838 (1999). 
\bibitem{27} Brenner, D. W., Phys. Rev. B 42, 9458-9471 (1990).
\bibitem{28} Kolmogorov, A. N. \& Crespi, V. H., Phys. Rev. B 71, 235415 (2005).

\end{thebibliography}
\end{document}